\documentclass[10pt]{aastex}
\usepackage{emulateapj5}
\usepackage{apjfonts}
\usepackage{psfig}

\newcommand{\kms}       {\mbox{km s$^{-1}$}}%
\newcommand{\kmsMpc}	{\mbox{km s$^{-1}$ Mpc$^{-1}$}}%

\shortauthors{Kannappan}
\shorttitle{Gas Fractions \& Galaxy Bimodalities}
\slugcomment{Accepted to ApJL}

\begin{document}

\title{Linking Gas Fractions to Bimodalities in Galaxy Properties}
\author{Sheila J. Kannappan\altaffilmark{1}}
\altaffiltext{1}{Harlan Smith Fellow, McDonald Observatory, The
	University of Texas at Austin, 1 University Station C1402,
	Austin, TX 78712-0259; sheila@astro.as.utexas.edu}

\begin{abstract} 

Galaxies over four decades in stellar mass are shown to obey a strong
correlation between $u-K$ colors and atomic-gas-to-stellar mass ratios
(G/S), using stellar mass-to-light ratios derived from optical colors. The
correlation holds for G/S ranging from nearly 10:1 to 1:100 for a sample
obtained by merging the SDSS DR2, 2MASS, and HYPERLEDA {H\,{\sc i}}
catalogs.  This result implies that $u-K$ colors can be calibrated to
provide ``photometric gas fractions'' for statistical applications.  Here
this technique is applied to a sample of $\sim$35,000 SDSS-2MASS galaxies
to examine the relationship of gas fractions to observed bimodalities in
galaxy properties as a function of color and stellar mass. The recently
identified transition in galaxy properties at stellar masses
$\sim2$--$3\times10^{10}$ M$_{\odot}$ corresponds to a shift in gas
richness, dividing low-mass late-type galaxies with G/S $\sim$ 1:1 from
high-mass galaxies with intermediate-to-low G/S.  Early-type galaxies below
the transition mass also show elevated G/S, consistent with formation
scenarios involving mergers of low-mass gas-rich systems and/or cold-mode
gas accretion.

\end{abstract}

\keywords{galaxies: evolution}

\section{\bfseries\sc\large Introduction}

Analyses of galaxies in the Sloan Digital Sky Survey (SDSS) have
demonstrated two distinct bimodalities in galaxy properties: a
bimodality between recent-burst dominated and more continuous star
formation histories (SFHs) as a function of stellar mass $M_{*}$,
divided at $M_{*}\sim3\times10^{10}$ M$_{\odot}$
\citep{kauffmann.heckman.ea:dependence}, and a bimodality between blue
late-type and red early-type galaxy sequences as a function of optical
color, divided at $u-r\sim2.2$
\citep{strateva.ivezi-c.ea:color,hogg.blanton.ea:luminosity,blanton.hogg.ea:broadband}.
Recently, \citet{baldry.glazebrook.ea:quantifying} have partially
unified these observations, demonstrating a color transition within
each of the two galaxy sequences at $M_{*}\sim2\times10^{10}$
M$_{\odot}$, as well as an increase in the relative number density of
red sequence galaxies above $\sim$2--5$\times10^{10}$ M$_{\odot}$.
They also argue that the number density of the red sequence is
consistent with a major-merger origin.  However, the cause of the
color and SFH transitions at $\sim$2--3$\times10^{10}$ M$_{\odot}$
remains to be explained.

Several physical processes that influence SFHs may imprint a
transition mass on the galaxy population.  Supernova-driven gas
blow-away will preferentially affect halos with small escape
velocities \citep{dekel.silk:origin}, although simulations suggest
that the baryonic mass threshold for blow-away may be closer to
$10^{7}$ M$_{\odot}$ than to $10^{10}$ M$_{\odot}$
\citep{mac-low.ferrara:starburst-driven}.  Cold-mode gas accretion may
dominate in low-mass halos whose gas fails to shock to the virial
temperature \citep{birnboim.dekel:virial,katz.keres.ea:how}; here
analytic estimates give a threshold mass of a few times $10^{11}$
M$_{\odot}$ including dark matter, so a link to the observed
transition at $M_{*}\sim$ 2--3 $\times10^{10}$ M$_{\odot}$ is
plausible.  Finally, observations suggest that inefficient star
formation may be typical of disk-dominated galaxies with $V_c \la 120
\kms$, possibly reflecting the relative importance of supernova
feedback as opposed to other turbulence drivers in supporting the
interstellar medium against gravitational instability
\citep{dalcanton.yoachim.ea:formation}.

All of these processes involve gas -- its expulsion, accretion, or
rate of consumption.  Thus examining how the gas properties of
galaxies vary with color and stellar mass may offer vital clues to the
origin of the transition mass and the color shifts within
the red and blue sequences.  Unfortunately, tracing the dominant
neutral phase of the interstellar medium requires {H\,{\sc i}} 21-cm line
observations, which are challenging even at the modest redshifts
probed by the SDSS.  To make full use of the statistical power
of the SDSS, an alternate strategy is required.

Building on earlier optical work \citep[e.g.,][]{roberts:integral},
\citet{bothun:searching} has shown a remarkably tight correlation
between {H\,{\sc i}} mass-to-$H$-band luminosity ratios and $B-H$ colors.
Going one step further, the present work describes a method for estimating
atomic-gas-to-stellar mass ratios using $u-K$ colors from the SDSS and
Two Micron All Sky Survey (2MASS) databases.  This ``photometric gas
fraction'' technique is calibrated using {H\,{\sc i}} data from the
recently expanded HYPERLEDA {H\,{\sc i}} catalog. When the technique is
applied to a sample of $\sim$35,000 SDSS-2MASS galaxies at $z<0.1$, the
transition mass of 2--3$\times10^{10}$ M$_{\odot}$ is observed to
correspond to a shift in gas richness found separately in both galaxy color
sequences.  This result implies that any explanation of the transition mass
via gas physics must directly or indirectly affect both early- and late-type galaxies.

\section{\bfseries\sc\large Data}

Optical, near-IR, and {H\,{\sc i}} data were obtained from the SDSS second
data release \citep[DR2,][]{abazajian:second}, the 2MASS all-sky extended
source catalog \citep[XSC,][]{jarrett.chester.ea:2mass}, and the HYPERLEDA
homogenized {H\,{\sc i}} catalog \citep{paturel.theureau.ea:hyperleda}.
Merged catalogs were constructed containing all $z<0.1$, $r<17.77$, $K<15$
galaxies with positions matched to within 6$\arcsec$ and with reliable
redshifts and magnitudes based on data flags and cataloged errors
(magnitude errors $<0.3$ in $K$, $<0.4$ in {H\,{\sc i}}, and $<0.15$ in
$ugr$).  The 2MASS magnitude limit was set fainter than the completeness
limit to improve statistics on dwarf and low surface brightness galaxies.
As the 2MASS XSC has uneven depth, it probes significantly fainter than the
completeness limit in some areas of the sky. Because of their marginal
detectability, galaxies with {H\,{\sc i}}-derived gas-to-stellar mass
ratios greater than two were targeted for individual inspection, and eight
were rejected as having unreliable 2MASS or SDSS pipeline reductions. These
rejections exacerbate the shortage of IR-faint galaxies.  The final samples
are: SDSS-HYPERLEDA (575 galaxies), SDSS-2MASS-HYPERLEDA (346 galaxies),
and SDSS-2MASS (35,166 galaxies).  An additional requirement for the
SDSS-2MASS sample was that the Local Group motion-corrected redshift be
greater than 1000 \kms.

All optical and IR magnitudes used here are fitted magnitudes, i.e.\ SDSS
model magnitudes and 2MASS extrapolated total magnitudes.  The SDSS
magnitudes are corrected for Galactic extinction using the DR2 tabulated
values and k-corrected to redshift zero using {\bf kcorrect v3.2}
\citep{blanton.brinkmann.ea:estimating}, while the 2MASS $K$-band
magnitudes are k-corrected using $k(z)=-2.1z$
\citep[][]{bell.mcintosh.ea:optical}.  Distances are computed in the
concordance cosmology $\Omega_{m}=0.3$, $\Omega_{\Lambda}=0.7$, $H_0=70$
\kmsMpc.

\section{\bfseries\sc\large Results}
\label{sc:results}

\begin{figure*}[t]
\epsscale{2.}
\plotone{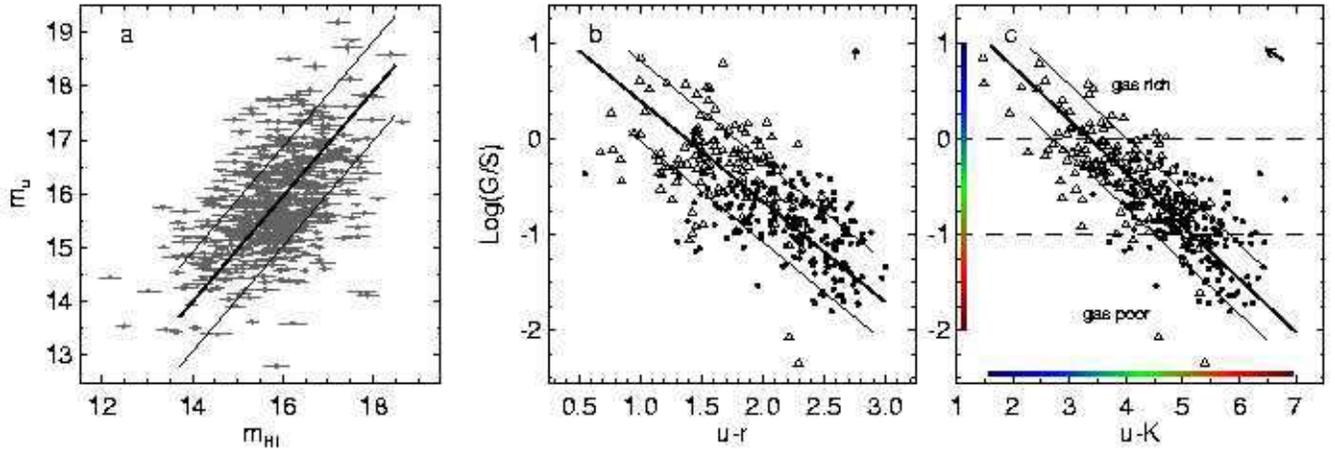}
\caption{{\em (a)} Apparent $u$-band magnitude vs.\ apparent {H\,{\sc
i}} magnitude for the SDSS-HYPERLEDA sample.  A bisector fit yields
$m_{u} = 0.33 + 0.98m_{\rm HI}$ (thick line) with $\sigma=0.92$ mag
(thin lines). {\em (b and c)} Atomic-gas-to-stellar mass ratio G/S
vs.\ $u-r$ and $u-K$ color for the SDSS-2MASS-HYPERLEDA sample.
Bisector fits yield $\log{(G/S)} = 1.46 - 1.06(u-r)$ and $\log{(G/S)}
= 1.87 - 0.56(u-K)$ (thick lines) with $\sigma=0.42$ dex and 0.37 dex,
respectively (thin lines).  Arrows indicate the effect of a 0.3 mag
error in $K$ (the maximum allowed by selection). {\em The color bar
in panel c provides a key to the photometric gas fractions in
Fig.~\ref{fg:colormass}.}  Horizontal lines demarcate gas-rich,
intermediate, and gas-poor regimes. Dots and triangles mark galaxies
with $M_{*}$ above and below $2\times10^{10}$ M$_{\odot}$,
respectively.}
\label{fg:both}
\end{figure*}

Fig.~\ref{fg:both}a shows the basic correlation between $u$-band and
21-cm apparent magnitudes $m_u$ and $m_{\rm HI}$ for the
SDSS-HYPERLEDA sample.  Its existence is not surprising: $u$-band
light is a tracer of young massive stars, and the birth rate of young
stars is known to depend on the available gas reservoir \citep[as in
the global correlation between disk-averaged star formation rate and
gas surface density,][]{kennicutt:star*1}.  The presence of young
massive stars may also enhance {H\,{\sc i}} detection
\citep[e.g.,][]{shaya.federman:h}.  The absolute magnitude correlation
obtained by distance-correcting $m_u$ and $m_{\rm HI}$ is of course
far stronger than the correlation in Fig.~\ref{fg:both}a, but at the
cost of non-independent axes.  In any case, what is relevant for
predicting $m_{\rm HI}$ from $m_u$ is not correlation strength but
scatter.  Most of the 0.92 mag scatter in the $m_u$--$m_{\rm HI}$
relation is not explained by the errors.  This scatter likely
represents variations in $u$-band extinction, molecular-to-atomic gas
ratios, and the physical conditions required to convert a gas
reservoir into young stars.  Even without calibrating these factors,
the $m_u$--$m_{\rm HI}$ relation is sufficiently tight for the present
application.

\begin{figure*}[t]
\epsscale{2.}  
\plotone{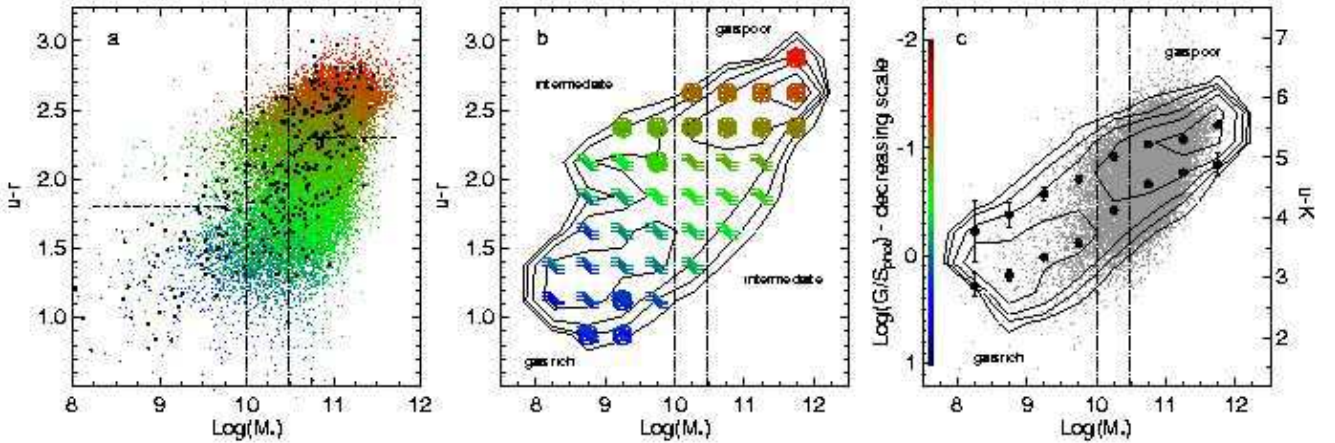}
\caption{Galaxy color vs.\ stellar mass for the SDSS-2MASS sample. Vertical
lines show the transition mass interval at 1--3 $\times$ 10$^{10}$
M$_{\odot}$.  {\em (a)} Individual $u-r$ data points color-coded by
photometric gas fraction G/S$_{phot}$ (computed using $u-K$). The dashed
line divides the red and blue sequences (referring to galaxy color, not
color-coding), approximating the optimal divider of
\citet{baldry.glazebrook.ea:quantifying}. The natural separation of the
sequences is clearest in the high-resolution online figure.  Black dots
show the {H\,{\sc i}} calibration sample of Figs.~\ref{fg:both}b and
c. {\em (b)} Gas fraction and morphology trends within the red and blue
sequences.  Contours show the conditional probability distribution of $u-r$
on $M_{*}$, in bins of $0.25\times0.5$ in color$\times$log($M_{*}$).  Each
galaxy is inverse-volume-weighted (see text) and all bins of a given mass
are normalized by the total for that mass column.  Each bin has three
slightly offset symbols whose colors show the weighted mean and
$\pm$1$\sigma$ values of G/S$_{phot}$, with errors propagated from the
scatter in the $u-K$ color--G/S relation.  Ovals and S-shapes indicate
early and late type bins based on weighted mean concentration index (see
text).  {\em (c)} G/S$_{phot}$ (or equivalently $u-K$) vs.\ stellar mass.
Small points show individual values, while contours give the conditional
probability distribution of G/S$_{phot}$ and $u-K$ on $M_{*}$, in bins of
$0.5\times0.5$ in color$\times$log($M_{*}$). Large points show weighted
mean G/S$_{phot}$ trends for the red and blue sequences separately, using
the $u-r$ division in panel a. All contours start at 0.04 and increase by
10$^{0.25}\times$ at each step.}
\label{fg:colormass}
\end{figure*}

Figs.~\ref{fg:both}b and~\ref{fg:both}c plot atomic-gas-to-stellar
mass ratios (G/S) against $u-r$ and $u-K$ colors for the
SDSS-2MASS-HYPERLEDA sample.  Gas masses are derived from {H\,{\sc i}}
fluxes with a helium correction factor of 1.4, and stellar masses are
derived from $K$-band fluxes using stellar mass-to-light (M/L) ratios
estimated from $g-r$ colors as in \citet{bell.mcintosh.ea:optical}.
The resulting correlations are distance-independent and extremely
strong, with Spearman rank correlation coefficients of 0.75 and 0.69
for $u-K$ and $u-r$ respectively.  Note that the calibration sample
spans the color--$M_{*}$ relation (Fig.~\ref{fg:colormass}a), and the
color--G/S relation is tighter than the color--$M_{*}$ relation for
these galaxies by $\sim$25\% in both $u-K$ and $u-r$.  The strength
of the $u-K$ color--G/S relation derives both from the underlying
$m_u$--$m_{\rm HI}$ relation and from the close correspondence between
$K$-band light and stellar mass.  The latter correspondence is assumed
within this work and may not apply to all starbursting systems
\citep{p-erez-gonz-alez.gil-de-paz.ea:stellar}; however,
\citet{kauffmann.heckman.ea:stellar} find that spectroscopically
determined M/L ratios generally agree well with color-based M/L
ratios, even in the low-mass regime where starbursts are common.

The large dynamic range of the $u-K$ color--G/S relation makes the
relation forgiving of errors and thus well-suited to low-precision
estimation of photometric gas fractions.  The 0.37 dex scatter in the
relation provides a basis for error estimation.  Furthermore, galaxies
of low and high mass define broadly similar $u-K$ color--G/S relations
in their regime of overlap (triangles and dots in
Fig.~\ref{fg:both}c).  It should be borne in mind that the generality
of the photometric gas fraction technique as currently formulated
relies on the fact that heavily dust-enshrouded star formation, as in
luminous infrared galaxies, is rare in the low-$z$ universe
\citep{sanders.mirabel:luminous}.  In dusty systems one might find
high G/S linked to {\em red} $u-K$ colors. Also, the calibration given
here could significantly underestimate actual gas-to-stellar mass
ratios if stellar M/L ratios are much lower than assumed and/or
molecular gas corrections are large \citep[a controversial topic;
see][]{casoli.sauty.ea:molecular,boselli.lequeux.ea:molecular}.  The
M/L ratios used here are roughly consistent with maximum disk
assumptions for spiral galaxies \citep{bell.mcintosh.ea:optical}.
Within the $ugriz$ magnitude set, the best alternative to the $u-K$
color--G/S relation is the $u-r$ color--G/S relation.  Its larger
scatter may in part reflect the fact that $K$-band magnitude errors
will move points along the $u-K$ color--G/S relation, but away from
the $u-r$ color--G/S relation.  However, the effect from cataloged
errors is quite small (arrows in Figs.~\ref{fg:both}b
and~\ref{fg:both}c), so the greater $u-r$ scatter seems to be mostly
physical.

Fig.~\ref{fg:colormass}a plots $u-r$ color vs.\ $M_*$ for the
$\sim$35,000-galaxy SDSS-2MASS sample, with points color-coded to indicate
photometric gas fractions G/S$_{phot}$ (computed from $u-K$ colors and the
fit in Fig.~\ref{fg:both}c).  The well-known red and blue sequences
(referring to galaxy color, not color-coding) are roughly separated by the
dashed line \citep[an approximation to the separator
of][]{baldry.glazebrook.ea:quantifying}.  Within each sequence, the
color-coding reveals a shift in gas fractions near a threshold mass of
$M_{*}^{t}\sim$ 1--3 $\times$ $10^{10}$ M$_{\odot}$.  Massive red-sequence
galaxies are extremely gas-poor (G/S$_{phot}$ as low as 1:100, red/yellow
color-coded points), whereas for red-sequence galaxies below $M_{*}^{t}$,
intermediate gas fractions (G/S$_{phot}$ $\sim$ 1:10, green points) are the
norm.  Likewise, massive blue-sequence galaxies have intermediate gas
fractions (G/S$_{phot}$ $\sim$ 1:10, green points), but blue-sequence
galaxies below $M_{*}^{t}$ are typically gas rich (G/S$_{phot}$ $\sim$ 1:1,
blue points).

These results are shown in binned form in Fig.~\ref{fg:colormass}b, where
the contours show the conditional probability distribution of $u-r$ on
$M_*$. In this plot, a vertical slice through the contours at a given mass
gives the one-dimensional probability distribution for $u-r$ colors at that
mass.  The conditional probability distribution is formed by weighting each
galaxy by 1/V$_{max}$, where V$_{max}$ is the maximum volume within which
it could have been detected, then normalizing the counts in each
color$\times$mass bin by the total counts in each mass column. Bins with
fewer than four galaxies are not considered. This algorithm is most robustly
applied to truly magnitude-limited samples, but the results shown here look
qualitatively similar to those of \citet{baldry.glazebrook.ea:quantifying}
for a magnitude-limited SDSS sample.  Each galaxy is assigned the
smallest of three V$_{max}$ estimates, based on (i) the magnitude limit in
$r$, (ii) the magnitude limit in $K$, and (iii) the distance limit imposed by
the $z<0.1$ selection requirement.

The color-coded symbols in Fig.~\ref{fg:colormass}b illustrate gas
fraction and morphology trends using 1/V$_{max}$-weighted bin
averages.  Ovals and S-shapes identify early and late type
morphologies, respectively, based on the concentration index $C_r$
(defined as $r_{90}/r_{50}$, where $r_{90}$ and $r_{50}$ are the 90\%
and 50\% Petrosian radii).  A value of $C_r=2.6$ is commonly used to
divide early and late types when true morphological type information
is lacking
\citep{strateva.ivezi-c.ea:color,kauffmann.heckman.ea:stellar,bell.mcintosh.ea:optical}.
Fig.~\ref{fg:colormass}b indicates transitional values
($2.55<C_r<2.65$) with both type symbols.  Some of the bluest low-mass
galaxy bins show transitional types, which will not be examined here.
Otherwise, the blue sequence forms a broad, low $C_r$ strip, with a
shift in gas richness at $M_{*}^{t}$ $\sim$1--3 $\times$ 10$^{10}$
M$_{\odot}$.  At a similar mass, the red sequence shows a coordinated
shift in gas richness and $C_r$.

Intriguingly, the increased gas content of the red sequence below
$M_{*}^{t}$ causes the red and blue sequences to fuse (nearly) in a plot of
G/S$_{phot}$ vs.\ $M_{*}$ (Fig.~\ref{fg:colormass}c).  The double-peaked,
slightly S-shaped contours of the G/S$_{phot}$ vs.\ $M_{*}$ distribution
thus look qualitatively similar to the bimodal distribution of SFH vs.\
$M_{*}$ reported by \citet{kauffmann.heckman.ea:dependence}.\footnote{Note
that the bimodalities in question are bimodalities in the conditional
probability distribution and need not appear directly in the observed
galaxy distribution.}  This similarity suggests that the bimodality in SFHs
may be intimately related to changes in G/S.  The contours in
Fig.~\ref{fg:colormass}c show a broad shift near $\sim$10$^{10}$
M$_{\odot}$, while the individual trends in the red and blue sequences
change slope over the range $\sim$1--3 $\times$ 10$^{10}$ M$_{\odot}$
(though the latter changes are not independent of the method of separating
the two sequences). Kauffmann et al.\ find a single-sequence bimodal
structure in $C_r$ vs.\ $M_{*}$ as well, perhaps because of the decreasing
$C_r$ of the red sequence below $M_{*}^{t}$.

Despite low $C_r$ values and high gas contents, the low-mass red
sequence does fit into the general merger scenario for the origin of
early-type galaxies, in a way that may illuminate the Kauffmann et
al.\ results.  The low-mass red sequence maps closely to an abundant
population of faint, moderately gas-rich S0 and S0/a galaxies observed
in the Nearby Field Galaxy Survey, a survey designed to represent the
natural distribution of galaxy types over a wide range of luminosities
\citep[][see Fig.~3 of the
latter]{jansen.franx.ea:surface,kannappan.fabricant.ea:physical}.
Based on the frequency of gas-stellar counterrotation in this
population and the scarcity of dwarf intermediate-type spiral
galaxies, \citet{kannappan.fabricant:broad} argue that $\ga$50\% of
low-luminosity S0's may form via late-type dwarf mergers.  Such
gas-rich mergers would naturally produce remnants with modest $C_r$,
explaining why low-luminosity early types are predominantly S0 rather
than E galaxies.  Moreover, low-luminosity S0's quite often have blue,
starbursting centers, despite red outer disks
\citep[][]{tully.verheijen.ea:ursa,jansen.franx.ea:surface}, and such
color gradients correlate with evidence of {\em recent} interactions
\citep[in all morphological types,][]{kannappan.jansen.ea:forming}. The SFH measures adopted by
\citet{kauffmann.heckman.ea:dependence}, which are based on
3$\arcsec$-aperture spectroscopy, may emphasize the starbursting
centers of this class of galaxies (and conversely, the quiescent
bulges of many high-mass late-type systems), reinforcing the
single-sequence structure of the resulting SFH vs.\ $M_{*}$ plots.

\section{\bfseries\sc\large Discussion}

A connection between star formation histories and gas fractions is in
some sense obvious: gas must be consumed to form stars, so old red
stellar populations will tend to be associated with diminished gas
supplies.  However, galaxies also accrete and expel gas, so this
simple view misses much of the story.  A complete picture must explain
why gas fractions depend on mass, and in particular why there is such
a close coincidence between the transition to gas richness
(atomic-gas-to-stellar mass ratio $\sim$ 1:1 in the blue sequence)
and the shift to recent-burst dominated SFHs below $\sim$
2--3 $\times$ 10$^{10}$ M$_{\odot}$.  Critical transition masses are
predicted by scenarios involving starburst-driven gas blow-away,
inefficient star formation below a gravitational instability
threshold, and/or cold-mode gas accretion
\citep{dekel.silk:origin,verde.oh.ea:abundance,birnboim.dekel:virial,katz.keres.ea:how,dalcanton.yoachim.ea:formation}.
The abundance of gas in low-mass (10$^8$--10$^{10}$ M$_{\odot}$)
galaxies seen in this paper is hard to explain if global gas blow-away
is a dominant process in this mass regime.  However, localized gas
blowout or strong feedback could inhibit efficient widespread star
formation in low-mass disk galaxies.  Such scenarios would not
explicitly account for the gas in low-mass red-sequence galaxies, but
formation via mergers of low-mass late-type systems, as discussed
above, could allow low-mass red-sequence galaxies to acquire their
modest gas excesses from gas-rich progenitors and thereby inherit the
progenitors' threshold mass for increased gas fractions.
Alternatively, it is possible that the transition in star formation
modes at $\sim$2--3 $\times$ 10$^{10}$ M$_{\odot}$ is not a cause, but
an {\em effect} of changing gas fractions, as in cold-mode accretion
scenarios.  If so, low-mass galaxies may form stars reasonably
efficiently and still appear gas rich.  The excess gas in low-mass
red-sequence galaxies could in this case represent post-merger
cold-mode accretion, possibly as part of a process of disk regrowth.

In conclusion, this paper has demonstrated a shift in galaxy gas mass
fractions at $M_{*}^{t}$ $\sim$ 1--3 $\times$ 10$^{10}$ M$_{\odot}$,
likely related to the shift in SFHs observed near the same stellar
mass by \citet{kauffmann.heckman.ea:dependence}.  The link may be
causal in either direction, depending on the relative importance of
supernova blowout, feedback, and cold-mode accretion processes in
determining $M_{*}^{t}$.  To establish this result, a technique has
been introduced to estimate photometric gas fractions based on the
correlation between $u-K$ colors and atomic-gas-to-stellar mass
ratios.  This correlation is interesting in its own right \citep[see
also][]{bothun:searching} and will be further examined and applied in
future work.

\acknowledgments

I thank D. Mar for being a helpful sounding board and S. Faber for sparking
my interest in galaxy bimodalities. N. Martimbeau and J. Huchra generously
shared a catalog merging code for me to modify.  I am also indebted to
S. Jester, C. Gerardy, D. Mar, and P. Hoeflich for assistance with database
and software issues.  K. Gebhardt and R. Kennicutt provided useful comments
on the original manuscript.  D. McIntosh, L. Matthews, A. Baker, and the
referee helped to improve the clarity of the final version.  This research
has used the HYPERLEDA {H\,{\sc i}} catalog (Vizier Catalog VII/238) and
data from 2MASS, a joint project of the U. of Massachusetts and
IPAC/Caltech, funded by NASA and the NSF. It has also used the SDSS (see
acknowledgement at http://www.sdss.org/collaboration/credits.html).  \\

\end{document}